\documentstyle[aps,epsf,prl,twocolumn]{revtex}
\begin{document}
\vspace{0.0cm}
\draft

\title{Quantum-Classical Correspondence for Isolated Systems of Interacting Particles: Localization and Ergodicity in Energy Space \thanks{presented to the Proceedings of the Nobel Simposia  ``Quantum Chaos Y2K"}}

\author{ F.M. Izrailev}
\address{Instituto de F\'isica, Universidad Aut\'onoma de Puebla,
Apartado Postal J-48, 72570, Puebla, M\'exico}
\maketitle

\begin{abstract}

Generic properties of the strength function (local density of states (LDOS))
and chaotic eigenstates are analyzed for isolated systems of interacting
particles. Both random matrix models and dynamical systems are considered in
the unique approach. Specific attention is paid to the quantum-classical
correspondence for the form of the LDOS and eigenstates, and to the
localization in the energy shell. New effect of the non-ergodicity of
individual eigenstates in a deep semiclassical limit is briefly discussed.

\end{abstract}


\section{Introduction}

In recent years, the growing attention was paid to the so-called {\it %
Quantum Chaos}, (see e.g. \cite{var00} and references therein). Nowadays,
this term is used in very different situations and often leads to a kind of
confusion. To clarify the subject, one needs to remind that the origin of
this term relates to quantum systems which, first, have the well-defined
classical limit, and second, in this limit the corresponding classical
system is assumed to manifest strong chaotic properties. One should stress
that systems under consideration are dynamical ones (or, the same, there are
no random parameters in their description). Therefore, this term was used in
relation to the problem of the quantum-classical correspondence for
dynamically chaotic systems.

Later, it was discovered that distinctive properties of {\it Quantum Chaos}
(properties of quantum systems with the dynamical chaos in the classical
limit) are generic for many other physical systems. For this reason, on
recent conferences and workshops on {\it Quantum Chaos} there were many
talks devoted to specific problems in atomic and nuclear physics, molecular
and solid state physics, optics and acoustics etc.

In order to classify the subject, we suggest to use global terms {\it %
Quantum Complexity} and {\it Wave Chaos}. The first term refers quantum
systems with complex behavior (or complex properties of spectrum and
eigenstates), both with and without the classical limit (see Fig.1).
Therefore, we stress that properties of complex quantum systems may have
generic features, although the mechanism can be different. In the case of 
{\it Quantum Chaos} the mechanism is closely related to the deterministic
classical chaos, in contrast to quantum systems without the classical limit,
where the mechanism of {\it complexity} is either due to a disorder ({\it %
Disordered Chaos}) or due to dynamical reasons (quite complex interactions,
although dynamical ones). As for the {\it Wave Chaos}, it refers to the
chaotic properties of classical systems described by wave equations which
are similar to the quantum Schr\"odinger equation. Therefore, many
properties of such classical systems have much in common with those of {\it %
Quantum Chaos} and {\it Quantum Disorder}. There are many physical
situations in electrodynamics, optics, acoustics, etc., where chaotic
properties of systems are well described by the methods developed in the
study of {\it Quantum Complexity}.

\vspace{-1.2cm}

\begin{figure}
\epsfxsize 7cm
\epsfbox{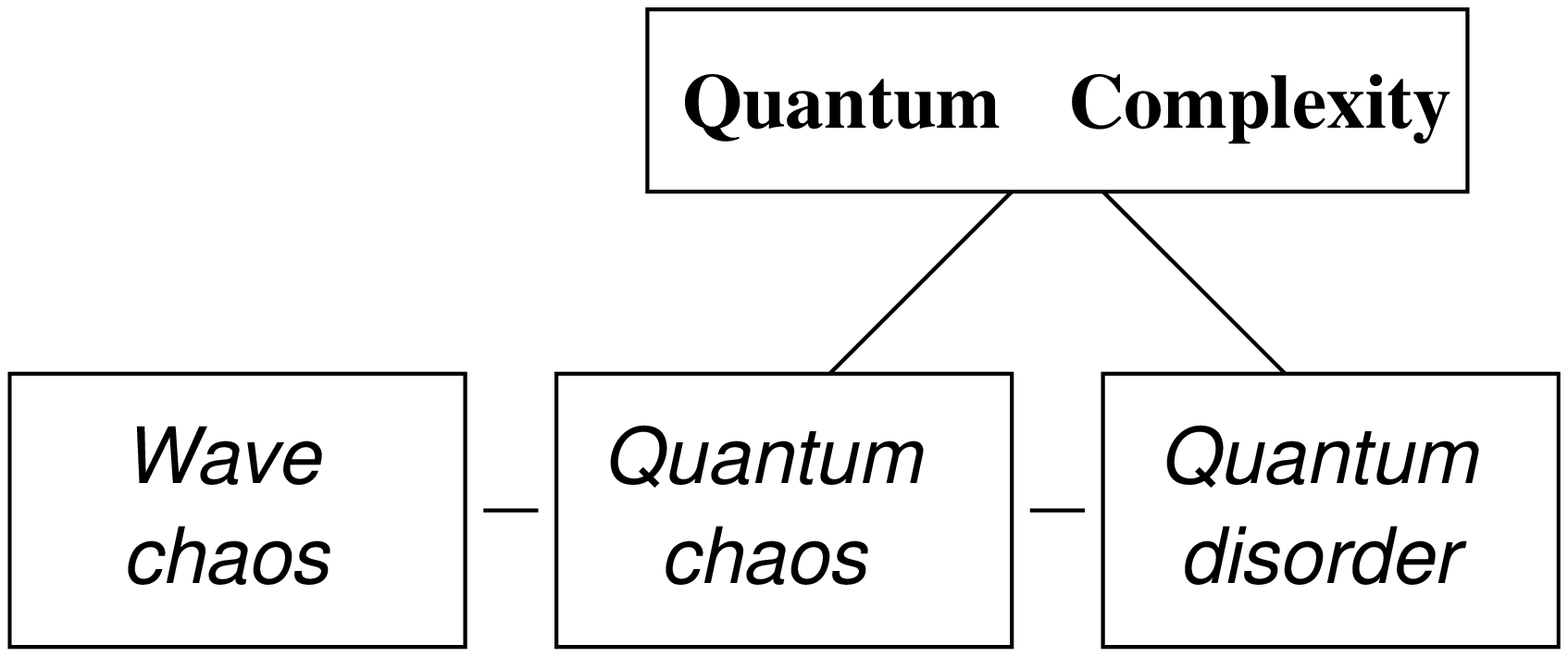}
\narrowtext
\vspace{-2.5cm}
\caption{}
\end{figure} 

Below we discuss generic features of complex quantum systems, both dynamical
and random ones, paying the main attention to the structure of strength
functions and chaotic eigenstates. In what follows we consider the models of
isolated systems of interacting particles which can be represented as a sum
of the unperturbed Hamiltonian $H_0$ and the interaction $V$, 
\begin{equation}
\label{H}H=H_0+V. 
\end{equation}
Here $H_0$ describes finite number of non-interacting particles or
quasi-particles. The term $V$ stands for a {\it two-body interaction}
between the particles, and it is assumed to be responsible for chaotic
properties of the total Hamiltonian $H$. For disordered models the
interaction $V$ is random by assumption, however, the two-body nature of an
interaction leads to important restrictions for chaos, see below. For
dynamical systems the mechanism of chaos is not obvious, however, it is
known to be related to classical chaos, if the classical limit exists. As
for dynamical quantum systems without classical limit, this case is less
studied, however, the knowledge of properties of {\it Quantum Chaos} and 
{\it Disordered Chaos} helps very much. We would like to stress that both
the Hamiltonian $H_0$ and interaction $V$ can be integrable (for dynamical
systems), however, the total Hamiltonian $H$ may reveal strong chaotic
properties.

Such a separation of a total Hamiltonian (\ref{H}) into two parts is common
in the description of complex atoms and nuclei and is known as the ''mean
field approximation''. The core of this approximation is to choose such a
basis in which regular part of the total Hamiltonian is absorbed in $H_0$
thus resulting in new single-particle states (quasi-particles). In contrast,
the {\it residual interaction} $V$ accumulates all other terms which can not
be embedded into $H_0$ due to their very complicated structure. We note that
the choice of the mean field is not well-defined procedure, however, one can
expect that if the most regular features of a system are described by $H_0$,
main results are not very sensitive to a specific choice of the mean field.

\section{Random matrix models}

In the limit case of a very strong and ''chaotic'' interaction $V\gg H_0$,
it is naturally to assume that system may be approximately represented by a
completely random matrix of a given symmetry, this idea was in the origin of
the {\it Random Matrix Theory} (see, for example, \cite{GMW99} and
references therein). Giving correct predictions for statistical fluctuations
of energy levels of complex physical systems, such as heavy nuclei and
highly excited many-electron atoms, full random matrices are too general and
can not describe global properties of physical systems such as dependencies
on the energy, strength of interaction, number of interacting particles etc.
It is interesting to note that one of the first random matrix models studied
in 1955 by Wigner (\cite{W55}) has more complicated form in comparison with
full random matrices. In analogy with (\ref{H}), it consists of two parts,
one of which is diagonal matrix with increasing entries $\epsilon_j$ and
another is a band matrix $V_{ij}$ with random elements inside the band, 
\begin{equation}
\label{wbrm}H_{ij}=\epsilon _j\delta _{ij}+V_{ij}. 
\end{equation}
In original papers \cite{W55} the ''unperturbed spectrum'' was taken in the
form of ''picked fence'', $\epsilon _j=jD$, where $D$ is the spacing between
two close energies and $j$ is a running integer number, however, more
generic case corresponds to random values $\epsilon _j$ with the mean
spacing $D$, reordered in an increasing way ($\delta_{ij}$ is the
delta-function). As for off-diagonal matrix elements $V_{ij}$, they are
assumed to be random and independent variables inside the band $|i-j|\le b$,
with the zero mean and given variance, $<V_{ij}=0>$ and $<V_{ij}^2>=V_0^2$.
Outside the band, the matrix elements equal to zero. Thus, the control
parameters of this model are the ratio $V_0/D$ of a typical matrix element
to the mean spacing, and the band-width $b$. As one can see, the first term
in (\ref{wbrm}) corresponds to the ``mean field'' $H_0$ and the fact that
the interaction has a finite energy range is directly taken into account.

In spite of impressive results of the RMT in many different applications
(see e.g. \cite{GMW99}), standard random matrix ensembles suffer from the
lack of information about the nature of interparticle interaction. For this
reason a new kind of random models was suggested in \cite{old} (see also the
review \cite{brody}); for a long time they were forgotten and recently they
have again attracted much attention in the connection with the onset of
chaos due to inter-particle interaction (see, for example, \cite{var99} and
references therein). Standard model of this kind describes $n$ interacting
particles which can occupy $m$ single-particle states ({\it orbitals}), for
Fermi-particles it has the form 
\begin{equation}
\label{tbre}H=\sum \epsilon _s\,a_s^{\dagger }a_s+\frac 12\sum
V_{s_1s_2s_3s_4}\,a_{s_1}^{\dagger }a_{s_2}^{\dagger }a_{s_3}a_{s_4}. 
\end{equation}
Here matrix elements $V_{s_1s_2s_3s_4}$ of the perturbation $V\,$stand for a 
{\it two-body interaction} (indices $s_1,s_2,s_3,s_4$ indicate initial $%
(s_1,s_3)$ and final $(s_2,s_4)$ single-particle states coupled by this
interaction), and $\epsilon _s$ is the energy of a single-particle state $%
\left| s\right\rangle $. Creation-annihilation operators $a_{s_j}^{\dagger }$
and $a_{s_j}$ define the many-particle basis $\left| k\right\rangle
=a_{s_1}^{\dagger }\,.\,\,.\,\,.\,a_{s_n}^{\dagger }\left| 0\right\rangle $
of non-interacting particles, in which the unperturbed Hamiltonian $H_0=\sum
\epsilon _s\,a_s^{\dagger }a_s$ is diagonal. It is convenient to reorder
this basis according to an increase of the unperturbed energy $%
E_k=\sum_s\epsilon _sn_s^{(k)}$ with an increase of the index $%
k=1,\,.\,.\,.\,,\,N\,$. In the {\it Two-Body Random Interaction} (TBRI)
model all matrix elements $V_{s_1s_2s_3s_4}$ are assumed to be random
independent variables. In realistic applications such as complex atoms and
nuclei, matrix elements of the two-body interaction are calculated directly
by using a proper mean field approximation, see, for example, \cite
{FGGK94,vict,zele}.

The TBRI model is very useful for the study of the role of a two-body
interaction, as well as for establishing generic properties of {\it Quantum
Complexity}, such as the localization, onset of chaos and thermalization in
closed systems. 
To compare with {\it Wigner Band Random Matrices} (WBRM) of the type (\ref
{wbrm}), one should stress the following. Having globally a band-like
structure, the matrix $H_{ij}$ defined by (\ref{tbre}) has many zero
off-diagonal matrix elements inside the band, even in the extreme case when
there are no additional integrals of motions and dynamical constraints,
therefore, when all single-particle states are coupled by the (two-body)
interaction. For $m\gg n\gg $ the ratio of the number of non-zero matrix
elements $V_{ij}$ to the total number of matrix elements is very small,
therefore, the matrix turns out to be very sparse. Moreover, non-zero
off-diagonal matrix elements $H_{ij}$ have been found \cite{FGI96} to be
slightly correlated, in spite of the complete randomness of two-body matrix
elements. The latter fact results in specific correlations between chaotic
compound states and may lead to non-statistical behavior of physical systems
with completely random two-body interaction even in the region where
conventional statistical description is assumed to be valid, see details in 
\cite{FGI96}.

The TBRI model (\ref{tbre}) allows both to study the conditions for the
onset of chaos and thermalization in closed systems, and to relate chaotic
properties of spectra and eigenstates in many-particle basis to the
properties of single-particle operators. One of important results obtained
recently \cite{AGKL97} in the frame of this model is the Anderson-like
transition which occurs in the Hilbert space determined by many-particle
states of $H_0$. The critical value $V_{cr}$ for this transition is
determined by the density of states $\rho_f=d_f^{-1}$ of those basis states
which are directly coupled be a two-body interaction. 
The point that the interaction should be compared not with the total density
of states $\rho _0$ of $H_0$ but with $\rho _f$ for the first time was noted
in \cite{aberg} when considering specific model of a rotating nuclei (see
also discussion in \cite{SS97}).

When the interaction is very weak, $V_0\ll d_f$, exact eigenstates are
delta-like functions in the unperturbed basis, with a very small admixture
of other components which can be found by the standard perturbation theory.
With an increase of the interaction, the {\it number of principal components}
$N_{pc}$ increases and can be very large, $N_{pc}\gg 1$. However, if the
interaction is still not too strong, $\pi ^{-2}\sqrt{d_fD}\ll V_0\ll d_f$ , 
\cite{FI97}\thinspace , the eigenstates are sparse, with extremely large
fluctuations of components. In order to have {\it ergodic eigenstates} which
fill some energy range (see below), one needs to have the perturbation large
enough, $V_0\gg d_f\,$(for a large number of particles this transition is
sharp and, in fact, one needs the weaker condition, $V_0\geq d_f$ ).

\section{LDOS: Breit-Wigner region}

In order to characterize global properties of ergodic eigenstates, it is
convenient to introduce the $F-${\it function} $F_j^{(n)}$ which gives the
envelope of eigenstates, 
\begin{equation}
\label{w}F_j^{(n)}\equiv \overline{w_j^{(n)}},\,\,\,\,\,\,\,\,
w_j^{(n)}\equiv \left| C_j^{(n)}\right| ^2=\left| C_j(E^{(n)})\right| ^2 
\end{equation}
Here $C_j^{(n)}$ are components of exact eigenstates $\left| n\right\rangle $
of the total Hamiltonian $H$ expressed in terms of basis states $\left| j
\right \rangle$ of $H_0$, $\left| n\right\rangle =\sum_jC_j^{(n)}\left|
j\right\rangle$, and the bar stands for the average inside small windows
centered at $j$ or $n$. We use here the notations which refer low indices to
the basis states, and upper indices to the exact (compound) eigenstates.
Thus, the structure of eigenstates is given by the dependence $w_j^{(n)}$ on 
$j$ for fixed values of $n$ .

On the other hand, if we fix the index $j$ and explore the dependence $%
w_j^{(n)}\,$ on $n$, one can find how the unperturbed state $\left|
j\right\rangle $ is coupled to other basis states due to interaction. The
latter quantity is very important since it gives the information about the
spread of the energy, initially concentrated in a specific basis state $%
\left| j\right\rangle \,,$ over other basis states, when switching on the
interaction. The envelope of this function $w_j^{(n)}\,$ in the energy
representation is related the {\it strength function } or {\it local
spectral density of states }(LDOS) which can be defined as follows, 
\begin{equation}
\label{LDOS}W(E^{(m)},j)=\sum_n|C_j^{(n)}|^2\delta (E-E^{(n)}) 
\end{equation}
The sum is taken over a number of eigenstates $\left| n\right\rangle \,$
chosen from a small energy window centered at the energy $E\,^{(m)}$ . One
can see that this function $W(E,j) $ is originated from the same matrix $%
w_j^{(n)}$ which gives the structure of eigenstates, therefore, the shapes
of strength functions and eigenstates are related to each other. Normalized
to the mean energy level spacing, the strength function $W(E,j)$ determines
an effective number $N_{pc}$ of principal components of compound states $%
\left| n\right\rangle $ which are present in the basis state $\left|
j\right\rangle $.

For the first time the form of the LDOS for WBRM (\ref{wbrm}) has been
analyzed in Ref. \cite{W55}. It was found that the form of the strength
function essentially depends on one parameter $q=\frac{\rho _0^2V_0^2}b$
only. Wigner analytically proved \cite{W55} that for relatively strong
perturbation, $V_0\gg D$, in the limit $q\ll 1$ the form of the LDOS is the
Lorentzian,

\begin{equation}
\label{BW} W_{BW}\,(\tilde E)= \frac 1{2\pi }\,\frac{\Gamma _{BW}}{\tilde
E^2+\frac {1}{4} \Gamma _{BW}^2},\,\,\,\,\,\,\,\,\tilde E=E-D\,j 
\end{equation}
which nowadays is known as the {\it Breit-Wigner} (BW) dependence. Here the
energy $\tilde E$ refers to the center of the distribution. The {\it %
spreading width} $\Gamma_{BW}$ (half-width of the distribution (\ref{BW}))
is given by the famous expression, 
\begin{equation}
\label{BWgam}\Gamma _{BW}=2\pi \rho _0V_0^2 
\end{equation}

In other limit $q\gg 1\,$ the influence of the unperturbed part $H_0\,$ can
be neglected and the shape of the LDOS\ tends to the shape of density of
states of band random matrix $V$, which is known to be the semicircle.

Recently, Wigner's results have been extended in \cite{FCIC96} to matrices $%
H $ with a general form of $V$, when the variance of off-diagonal matrix
elements decreases smoothly with the distance $r=\left| i-j\right| $ from
the principal diagonal. In this case the effective band size $b\,$ is
defined by the second moment of the envelope function $f(r)$ . Another
important generalization of the WBRM studied in \cite{FCIC96}, is an
additional sparsity of the matrix $V$, which may mimic realistic
Hamiltonians.

Random matrix models of the type (\ref{wbrm}) are very useful for the
understanding some important properties of the LDOS. Let us, first, write
the condition for the LDOS to be of the BW form \cite{FCIC96}, 
\begin{equation}
\label{range1}D \ll \,\Gamma _{BW}\ll \Delta_E;\,\,\,\,\,\, \Delta_E=b D 
\end{equation}
The left part of this relation refers to the non-perturbative character of
the interaction, according to which many of unperturbed basis states are
strongly coupled by an interaction. On the other hand, the interaction
should be not very strong, namely, the width $\Gamma_{BW}$ determined by Eq.(%
\ref{BWgam}), has to be less than the width $\Delta_E$ of the interaction in
energy representation. The latter condition is generic for systems with
finite range of the interaction $V$. One should also stress that, strictly
speaking, the BW form is not correct for such systems since its second
moment diverges. As was shown in \cite{FGGK94}, outside the energy range $%
|\tilde E|> \Delta_E$ the LDOS in the model (\ref{wbrm}) decreases faster
than pure exponent.

For other models, such as the TBRI model where the band width $b$ of the
interaction is not well-defined, instead of $\Delta _E$ it is more
convenient to use another important quantity, the variance $\sigma _0^2$ of
the LDOS. The latter quantity can be rigorously expressed through
off-diagonal matrix elements of the interaction, $\sigma _0^2=\sum_jV_{ij}^2$
for $i\neq j$, therefore, $\sigma _0^2=2bV_0^2$ for the model (\ref{wbrm}).
As a result, we have $\Delta _E=\pi \sigma _0^2/\Gamma _{BW}$ and Eq.(\ref
{range1}) can be written as 
\begin{equation}
\label{range2}D\ll \,\Gamma _{BW}\ll \sigma _0\sqrt{\pi } 
\end{equation}
Numerical data \cite{FCIC96,FI97,CFI00} for the WBRM and TBRI models show
that on the border $\Gamma _{BW}\approx 2\sigma _0$ the form of the LDOS is
quite close to the Gaussian, and this transition from the BW dependence to
the Gaussian-like turns out to be quite sharp. One should note that the form
of the LDOS determines the dynamics of wave packets in the energy
representation, see the data for the WBRM in Ref.\cite{CIK00}.

\section{LDOS: transition to the Gaussian}

For a long time it was assumed that in real physical systems the LDOS has
the universal BW-dependence (\ref{BW}). However, when studying the structure
of the LDOS\ and eigenfunctions of the Ce atom, it was observed clear
deviation from the BW-shape. Moreover, recent numerical investigation of
nuclear shell models \cite{zele} has shown that the form of the LDOS is much
closer to the Gaussian rather that to the BW. This fact is due to a quite
strong interaction $V\sim H_0\,,$ since the mean field $H_0$ often includes
a large regular part of the interaction, thus leaving a ``disordered part'' (%
{\it residual interaction}$)$ in $V$ .

As one can see, it is of great importance to find an analytical description
of the LDOS in dependence of the strength of interaction. Although the
extreme limit of a very strong interaction, $q\gg 1,$ (or, the same, $W_{BW}
\gg \sigma_0$) has been studied by Wigner in the WBRM model (\ref{wbrm}) ,
the semicircle form of the LDOS seems to be unphysical. Indeed, this form is
originated from the semicircle dependence of the total density $\rho _V(E)$
, and the latter is known to be an artifact of the standard RMT. As was
shown quite long ago (see the review \cite{brody} and reference therein),
this result formally corresponds to random $n-$body interaction between
Fermi or Bose particles with $n=1,...\infty .$ On the other hand, physical
interaction is typically of a two-body nature, this fact is directly taken
into account in TBRI models of the type (\ref{tbre}).

It is clear that analytical treatment of the TBRI models is much more
difficult in comparison with full and band random matrices. For this reason,
numerical data are very important since they may give a hint for rigorous
results to be proved analytically (see, for example, \cite{FGI96}). The form
of the LDOS\ in the TBRI model (\ref{tbre}) in the extreme case when $H_0$
can be neglected, is also defined by the density of states $\rho _V$
associated with the random two-body interaction $V$ . As was shown in \cite
{old}, for large number of particles and orbitals, $m\gg n\gg 1$ , the
density of states $\rho _V$ has the gaussian form. This result is expected
from the point of view of combinatorics, however, the rigorous proof is
non-trivial.

Using general result of the WBRM model, see Eq.(\ref{range2}), one can
expect that if the spreading width $\Gamma _{BW}\,$ , determined by the
density $\rho _f=d_f^{-1}$ of directly coupled basis states, larger than the
``width'' $2\sigma _0$ of the LDOS, 
\begin{equation}
\label{lam}\lambda \equiv \frac{\Gamma _{BW}}{2\sigma _0}\geq
1\,,\,\,\,\,\,\,\Gamma _{BW}=2\pi \rho
_fV_0^2\,,\,\,\,\,\,\,V_0^2=\left\langle V_{ij}^2\right\rangle \,, 
\end{equation}
the form of the LDOS approaches the Gaussian. As is indicated above, the
variance $\sigma _0^2$ of the LDOS is defined by the off-diagonal matrix
elements of $V_{i\neq j}$ and can be found analytically (see details in \cite
{FIC96,FI97}), 
\begin{equation}
\label{sigma}\left( \sigma _0^2\right) _j=\sum_{i=1}^NV_{ij}^2=\frac{V_0^2}%
4n(n-1)(m-n)(m-n+3)\,\,. 
\end{equation}
It turns out that for Fermi-particles the variance of the LDOS is
independent of the index $j$ (therefore, of the total energy $E^{(j)}$ ), in
contrast to the models with Bose-particles, for this reason in what follows
we omit the index $j\,$ for $\sigma _0$ . One can see that with an increase
of the interaction, the half-width $\Gamma _{hw}$ of the LDOS changes from
the quadratic dependence $\Gamma _{hw}\approx \Gamma _{BW}\sim V_0^2$ to the
linear one, $\Gamma _{hw} \sim V_0$. Sometime, this fact is missing in the
analysis of the TBRI models. It is important to stress that in the case of $%
\lambda \geq 1$ the parameter $\Gamma _{BW}$ has nothing to do with the
half-width $\Gamma _{hw}$ of the LDOS, the latter is proportional to $\sigma
_0$ (see also discussion in \cite{zele}).

Very recently, the form of the LDOS for the TBRI model (\ref{tbre}) was
analytically found \cite{FI00} in general form. Since the density of states $%
\rho (E)$ strongly depends on the energy, one should take it into account
explicitly, $W_k(E)=F(E_k,E)\,\rho (E)$ . Here, $F(E_k,E)\,$ is the $F-$%
function discussed above, written in the energy representation, $E\equiv
E^{(i)}\,$ is the total energy (energy of an exact state $\left|
i\right\rangle $ of $H$ ), and $E_k$ is the unperturbed energy (energy of
the basis state $\left| k\right\rangle $ of $H_0$ . The method used in \cite
{FI00}, is an extension of the approach developed in \cite{BM69}, which
takes into account specific structure of the Hamiltonian $H$ . The
dependence of $W_k(E)$ was found to have the form,

\begin{equation}
\label{FfBW}W_k(E)=\frac 1{2\pi }\frac{\Gamma _k(E)}{(E_k-E)^2+\frac
14\,\Gamma _k^2(E)} 
\end{equation}
with some function $\Gamma _k(E)$. In the case of a relatively weak
interaction, $1\ll \lambda \leq 1$ , the function $\Gamma _k(E)$ is almost
constant on the scale of the energy width $\Delta _E$ , therefore, the
conventional BW-expression (\ref{BW}) is recovered for $W_k(E)$ . In
contrast, in the case of strong interaction, $\lambda \geq 1$ , the
dependence of $\Gamma _k(E)\,$ on the energy $E$ can not be neglected and
the nominator in (\ref{FfBW}) is the leading one. As was shown in \cite{FI00}%
, the function $\Gamma _k(E)$ is defined by the density of states for
one-body and two-body transitions, and for large number of particles and
orbitals has the Gaussian form with the variance $\sigma _0$ . Detailed
numerical study \cite{CFI00} of the form of the LDOS in the region $\Gamma
_{BW}\geq 2\sigma _0$ have shown that the LDOS\ coincides with the Gaussian
with a very high accuracy.

It is important to stress that the condition for the existence of a smooth
energy dependence $\Gamma_k(E)$ is defined by the condition $V_0\geq d_f$ of
ergodicity for the components $C_k^{(i)}$ of the LDOS (or, the same, for
exact eigenstates). Below this transition, for $V_0\leq d_f$ , the smooth
solution of the equation for the function $\Gamma _k(E)$ does not exist. In
this sense, the result obtained in \cite{FI00} can be treated as the
independent proof of the onset of chaos in the TBRI model when $V_0\geq d_f$
.

Having formal solution (\ref{FfBW}) for the LDOS, it is, however, convenient
to find a simple phenomenological expression which depends on two control
parameters only, the width $\Gamma _{BW}$ defined by (\ref{lam}), and the
width $\sigma _0$ , see Eq.(\ref{sigma}). This was recently done in \cite
{CFI00} where both the WBRM and TBRI models have been used to compare
numerical data with the following expression, 
\begin{equation}
\label{fit}W(z;\lambda ,\beta )=A\frac{\exp \left( -\gamma ^2\frac{z^2}%
2\right) }{z^2+\frac{\beta ^2}4\Gamma _{BW}^2} 
\end{equation}
Here $z=(E-E_c)/\sigma _0$ is the normalized energy which refers to the
center of the LDOS, with $\sigma _0$ given by (\ref{sigma}). Taking the
parameter $\beta $ as the fitting parameter, two other parameters $A$ and $%
\gamma $ are determined from the normalization conditions $\int W(z)\,dz\,=1$
and $\int z^2W(z)\,dz\,=1$ . As one can see, in the normalized energy units
there is only one control parameter $\lambda $ . In the region $\lambda \ll 1
$ the BW-dependence for $W(z)$ is recovered, and for $\lambda \gg 1$ the
Gaussian emerges, see details in \cite{CFI00}.

Numerical data show that this dependence gives quite good description of the
LDOS for any values of $\lambda $ defined by $\Gamma _{BW}$ and $\sigma _0$
. It turns out that the transition from the BW-dependence to the Gaussian is
quite sharp and takes place when $\lambda \approx 1.0$ (see also recent
numerical data \cite{kota}). The above phenomenological expression (\ref{fit}%
) with the fitting function $\beta ($$\lambda )$ is quite useful in the
applications.

So far, we have discussed the form of the LDOS, provided the eigenstates are
chaotic in the sense that the fluctuations of their components $C_j^{(i)}$
around the smooth envelope given by $W(E)$ are gaussian-like. Similar
analysis can be performed for the form $w(E)\,$of exact compound eigenstates
in the unperturbed energy representation. As was noted above, these two
functions, $W(E)$ and $w(E)$ are related to each other through the matrix $%
w_j^{(n)}$ , see Eq.(\ref{w}). Analytical estimates \cite{FI97} for the TBRI
model show that for not very strong interaction, when $\sigma _0\ll \Delta E$
with $\Delta E$ as the total width of the energy spectrum of $H_0$ , the two
functions are close to each other. This fact allows to use the results
obtained for the LDOS, when considering typical structure of exact
eigenstates.

\section{Distribution of occupation numbers}

The knowledge of the $F-$function is very important for the relation of
chaotic properties of the Hamiltonian $H$ in the many--body representation,
with the properties of single-particle operators. Indeed, exact eigenstates $%
\left| i\right\rangle \,$ of the total Hamiltonian $H$ are given as

{\it 
\begin{equation}
\label{basis}\left| i\right\rangle =\sum_kC_k^{(i)}\left| k\right\rangle
,\,\,\,\,\,\,\,\,\,\,\,\left| k\right\rangle =a_{s_1}^{\dagger
}\,.\,\,.\,\,.\,a_{s_n}^{\dagger }\left| 0\right\rangle 
\end{equation}
}where {\it $C_k^{(i)}$} is the $k-th$ component of the compound state {\it $%
\left| i\right\rangle $} in the unperturbed basis. These coefficients $%
C_k^{(k)}$ determine the so-called {\it occupation numbers} $n_s$ ,

\begin{equation}
\label{ns}n_s^{(i)}=\left\langle i\right| \hat n_s\left| i\right\rangle
=\sum_k\left| C_k^{(i)}\right| ^2\left\langle k\right| \hat n_s\left|
k\right\rangle 
\end{equation}
where $\hat n_s=a_s^{\dagger }a_s$ stands for the occupation number
operator. This quantity gives the probability that one of $n$ particles
occupies an orbital $s$ specified by the one-particle state $\left|
s\right\rangle $  for the fixed exact  state $\left|
i\right\rangle \,$ . According to this expression, this probability can be
found by projecting the state $\left| i\right\rangle \,$ onto the basis of
unperturbed states, for which the relation between the positions of all
particles in the single-particle basis and the specific many-particle basis
state is known by the construction of the latter. One can see that the
probability $n_s=n_s(E^{(i)})$ is the sum of probabilities over a number of
basis states which construct the exact state. For Fermi-particles the
occupation number $n_s^{(k)}=\left\langle k\right| \hat n_s\left|
k\right\rangle $ is equal to $0\,$or $1$ depending on whether any of the
particles in the basis state $\left| k\right\rangle $ occupies or not the
single-particle state $\left| s\right\rangle $ .

For chaotic eigenstates the $n_s-$distribution is a fluctuating function of
the total energy $E=E^{(i)}$ of a system, due to strong (gaussian)
fluctuations of the components $C_k^{(i)}$. In order to
obtain a smooth dependence, one should make an average over a small energy
window centered at $E^{(i)}$ , which is in the spirit of the conventional
statistical mechanics for systems in the contact with thermostat. In fact,
such an average is a kind of microcanonical averaging since it is done for
the fixed total energy $E$ of a system. Therefore, one can define the $n_s-$%
{\it distribution} through the $F-function,$ see (\ref{w}) and (\ref{ns}), 
\begin{equation}
\label{nsAv}n_s(E)=\sum\limits_kF(E_k,E^{(i)}) \, \left\langle k\right| \hat
n_s\left| k\right\rangle . 
\end{equation}

The $n_s-$ distribution plays essential role in the statistical approach to
finite systems of interaction particles, see details in \cite{FI97,var99}.
It is clear that non-trivial part in the above expression for $n_s$ is the $%
F-$function $F(E_k,E^{(i)})$ which absorbs statistical effects of the
two-body interaction $V$. The important point is that in order to find the
distribution of occupation numbers, one needs to know the envelope of exact
eigenstates in the basis (or energy) representation, not the eigenstates
themselves. Therefore, the knowledge of the $F-$function and the density of
states $\rho _0(E)$ of the unperturbed Hamiltonian $H_0$ (it appears when
passing from the summation to the integration in Eq.(\ref{nsAv})) gives us
the possibility to relate global properties of eigenstates to the
distribution of occupation numbers of single--particle states.

The form of $n_s\,$ in the TBRI\ model (\ref{tbre}) has been studied in
detail in connection with the problem of thermalization and of the onset of
Fermi-Dirac (FD) distribution. It should be stressed that the occurrence of
the FD-distribution in closed systems of finite number of interacting
particles is far from being trivial, especially, when the number $n$ of
particles is not very large. It has to be noted that in many applications $n$
is the number of particles (quasi-particles) above the Fermi level,
therefore, it can be relatively small. For example, in the mean field
description of the Ce atom there are $n=4$ interacting electrons in the
outer shell \cite{FGGK94,vict}, and $n=12$ in the standard $s-d$ shell model
of heavy nuclei \cite{zele}.

One of the most interesting results obtained in the frame of the TBRI model (%
\ref{tbre}) \thinspace is that the FD-distribution may occur even for 4
interacting Fermi-particles, provided the (random) interaction $V\,$is
strong enough. In this case one can obtain an analytical expression for the
temperature and chemical potential which stand in the standard Fermi-Dirac
form for the occupation number distribution $n_s$. Another important result
is that even if the form of the $n_s-$distribution is far from the
FD-distribution, it can be described analytically, the simplest case is when
the form of the $F-$function is close to the Gaussian. The study of the $n_s-
$distribution allows to determine the onset of thermalization in terms of
the ergodicity of components of compound states in the energy range defined
by the $F-$function. In fact, the ergodicity of eigenstates results in
standard (gaussian-like) fluctuations, both for components of eigenstates
around the envelope ($F-$function) and for the $n_s$ numbers, when changing
slightly the total energy of a system (see details in \cite{FIC96,FI97,var99}%
). The latter property (existence of a smooth dependence of the $n_s-$%
distribution with gaussian fluctuations around this dependence) is , in
essence, the existence of a statistical equilibrium for interacting
particles in a closed system.

\section{Quantum-classical correspondence}

One of the important questions is the quantum-classical correspondence for
the {\it shape of eigenstates} (SE) and the LDOS. As was pointed out in Ref.%
\cite{CCGI96}, there is quite simple approach for finding {\it classical
shape of eigenstates} and {\it classical LDOS.} This approach is generic and
can be used for different physical models \cite{BGIC98,BGI98,we}.

Let us start with the classical SE. We consider the total Hamiltonian in the
form (\ref{H}) where the ``unperturbed part'' $H_0$ can be represented as
the sum of single-particles Hamiltonians $H_k^0$ describing the motion of $n$
non-interacting particles, $H_0=\sum_{k=1}^nH_k^{\,0}(p_k,q_k)$. The
interaction $V$ \thinspace between the particles is assumed to result in a
chaotic behavior of the system $H$. Now let us fix the {\it total} energy $%
E\,$ of the Hamiltonian $H(t)$ and find (numerically) the trajectory $%
p_k(t)\,,\,q_k(t)$ by computing Hamiltonian equations. Since the total
Hamiltonian $H$ is chaotic, any choice of initial conditions $%
p_k(0)\,,\,q_k(0)$ gives the same result if one computes the trajectory for
a sufficiently large time. When time is running, let us collect the values
of the {\it unperturbed }Hamiltonian $H_0(t)$ at fixed times $%
t=T,2T,3T,\,...\,,$ and construct the distribution of energies $E_0(t) $
along chaotic trajectory of the {\it total \thinspace }Hamiltonian $H$ . In
this way one can obtain the distribution $w(E_0;E=const)\,\,$ . Comparing
with the quantum model, one can see that this function $w(E_0;E=const)\,$ is
the classical analog of the average shape of eigenstates in the energy
representation. Indeed , any exact eigenstate corresponds to a fixed total
energy $E=const$ and it is represented in the unperturbed basis of $H_0$ .
Thus, one can expect that for chaotic eigenstates in a deep semiclassical
region the two above quantities, classical and quantum ones, correspond to
each other.

In the same way one can consider the complimentary situation. Let us fix the 
{\it unperturbed} energy $E_0$ and compute the trajectory $%
p_k^{(0)}(t)\,,\,q_k^{(0)}(t)$ which belongs to the {\it unperturbed}
Hamiltonian $H_0(t)\,.\,$ Similar to the previous case, let us put this
unperturbed trajectory into the total Hamiltonian $H(t)$ and collect the
values of the total energy $E(t)$ along the unperturbed trajectory for
discrete values of time. It allows us to find the distribution $%
W(E;E_0=const)$ which now should be compared with the LDOS in the
corresponding quantum model. However, in this case one should perform an
average over many initial conditions $p_k(0)\,,\,q_k(0)$ with the same
energy $E_{0\text{ }}$ , since the unperturbed Hamiltonian $H_0$ is
separable. One can see that the above two classical distributions $%
w(E_0;E=const)$ and $W(E;E_0=const)$ determine the ergodic measure of two
energy shells, the first one, when projecting the phase space surface of $H\,
$ onto $H_0$ , and the second one, when projecting the surface of $H_0\,$
onto $H$ (see discussion in \cite{CCGI96}).

\section{Two interacting spins}

The first example is the model of two interacting rotors (see \cite
{BGIC98,BGI98} and references therein),

\begin{equation}
\label{ham1}H=L_z+M_z+L_xM_x 
\end{equation}
with angular momentum $\vec L$ and $\vec M$ . This model may be used to
describe the interaction of quasi-spins in nuclear physics or pseudo--spins
in solid state systems. Comparing with Eq.(\ref{H}), here $H_0=L_z+M_z$ and $%
V=L_xM_x$. The constants of motion are $H=E$, $L^2$ and $M^2\,$, which are
connected by the relation $E^2\leq E_{max}^2=(L^2+1)(M^2+1) $ for $LM>1$. It
is worth to mention that in (\ref{ham1}) the dynamical variables $\vec
L,\vec M$ are not canonical, however, keeping $L^2$ and $M^2$ as constants,
one can present $H\,$ in the canonical variables.

It is known (see references in \cite{BGIC98,BGI98}) that in the classical
limit the phase space consists of both regular and chaotic regions, the size
of which depends on the values of $L^2$ $\,$ and $M^2\,$. We consider the
simplest case $L=M\,$ for which trajectories are regular when $|E|$ is close
to $E_{max}=L^2+1$ , and are chaotic at the center of the energy band, $%
E\simeq 0 $ .

Quantization of this model can be done in the standard way, with angular
momenta quantized according to the relations $L^2=M^2=\hbar ^2l(l+1)$ where $%
l$ is an integer number. Therefore, for a given $l$ the Hamiltonian matrix
is finite and the semiclassical limit corresponds to the limit $l\to \infty $
and $\hbar \to 0$ while keeping $L^2$ constant. According to the approach
discussed above, the Hamiltonian is represented in the two--particles basis $%
|l_z,m_z\rangle $ and has the dimension $N=2l^2+2l+1$. Due to the symmetry
degeneracy with respect to the exchange of particles, there are symmetric
and antisymmetric states which are not coupled by the interaction (only
symmetric states are studied in \cite{BGIC98,BGI98}).

As was pointed out when discussing the TBRI matrix model (\ref{tbre}), it is
important to reorder the unperturbed basis in such a way that the energy of
many-particles states of $H_0\,$ increases. This kind of sorting corresponds
to the classical model and allows to establish quantum-classical
correspondence for the shape of LDOS and eigenstates. After this reordering,
the unperturbed Hamiltonian $H_0$ has the shell structure, with the same
value on energy inside one shell. Diagonal matrix elements are given by the
eigenvalues $-2l\hbar ,(-2l+2)\hbar ,\ldots ,2l\hbar $ of the unperturbed
Hamiltonian $H_0$, and diagonal elements of the perturbation $V$ vanish due
to a particular form of the interaction.

The global structure of the matrix $H_{m,n}$ is quite specific,  however, it
has some similarity with the TBRI matrices. Apart from the principal
diagonal $H_{ii}$ determined by the eigenvalues of $H_0$ , the next (to the
principal one) diagonals $H_{n,n\pm 1}$ correspond to transitions inside
each $H_0$-shell while the ''arcs'' connecting the two corners represent
transitions between neighboring shells having $\Delta H_0=\pm 2\hbar $.
Therefore, the matrix $H_{ij}$ has only four off-diagonal ``curves'' along
which non-zero off-diagonal elements are located. All other matrix elements
vanish thus resulting in extremely sparse matrix $H_{ij}\,.$ In spite of a
clear band structure, the Hamiltonian matrix $H_{ij}$ can not be compared
with Wigner Band Random Matrices of the type (\ref{wbrm}) since its band
size is not much less than the total size of the matrix. Indeed, the band
width is $b=2l+1$ in the middle of the matrix, therefore, the control
parameter $b^2/N$ is of the order 1 and the finiteness of the matrix is
crucial (see details in \cite{BGIC98,BGI98}). It is also important to stress
that non-zero matrix elements can not be treated as random ones since the
model (\ref{ham1}) is dynamical. All non-zero off-diagonal matrix elements
are positive and the mean and variance of the distribution of these matrix
elements depend on the classical parameter $L^2=\hbar ^2l(l+1)$ only.

Numerical data for the LDOS in the center of the energy band , $E\approx 0$
, have been obtained in the dynamical approach discussed in the previous
Section. More precisely, this distribution was numerically calculated taking
a sample of chaotic trajectories having the same fixed values of $E$ and $%
L^2=M^2$. Following these trajectories, one can calculate $%
H_0(t)=L_z(t)+M_z(t)$ taken at equal instants of time and find the
distribution of $H_0$. In this case the classical distribution can be also
evaluated analytically since the classical unperturbed Hamiltonian $H_0$ is
integrable.

The result is presented in Fig.2 in the normal and semi-log scales. The
quantum distribution Eq.(\ref{LDOS}) was obtained by averaging over $l+1$
values of individual dependencies $w_l^{(k)}$ for the largest shell with $%
\epsilon_0=0$ . Here, the normalized energies are used, $\epsilon
_0=E_0/E_{\max }$ and $\epsilon =E/E_{\max }$ , with $E_0$ and $E$ as the
unperturbed and exact energies. One can see a remarkable correspondence
between the quantum LDOS\ and its classical counterpart. The only difference
can be found in long tails of the distribution, which manifests a kind of
quantum tunneling beyond the energy region given by the classical
Hamiltonian. Also, one can notice some discrepancy at the center of the
energy band, where quantum effects are also expected to be relatively strong
due to a finite value of the semiclassical parameter $l$ .

A very good correspondence has been also found for the shape of exact eigenstates SE, see Fig.3. Numerical procedure of the averaging was taken in the way similar to that for the LDOS, with the average over $l+1$ exact eigenstates taken from the center of the energy band. Again, one can see some difference in the tails and at the center of the distribution, which is caused by pure quantum effects.

\vspace{-2.3cm}
\begin{figure}
\hspace{.8cm}
\epsfxsize 7cm
\epsfbox{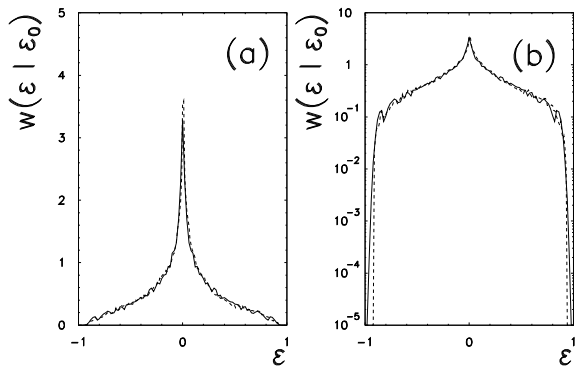}
\vspace{-2.5cm}
\narrowtext
\caption{ The LDOS distribution (full line) and
its classical counterpart (dashed line)
for the case $L=3.5$, $l=39$, after [24].}
\label{figLDOS}
\end{figure} 

It should be pointed out that at the edges of the energy spectrum where the
classical motion is regular, individual eigenstates differ very strongly one
from the other, therefore, one needs take a relatively large number of
eigenstates for the average, in order to obtain a smooth dependence on the
energy. This means that although the quantum-classical correspondence can be
achieved in the regular region, it has no sense since the energy interval to
be taken for the average can be of the order the whole spectrum.

\vspace {-2.3cm}
\begin{figure}
\hspace{.8cm}
\epsfxsize 7cm
\epsfbox{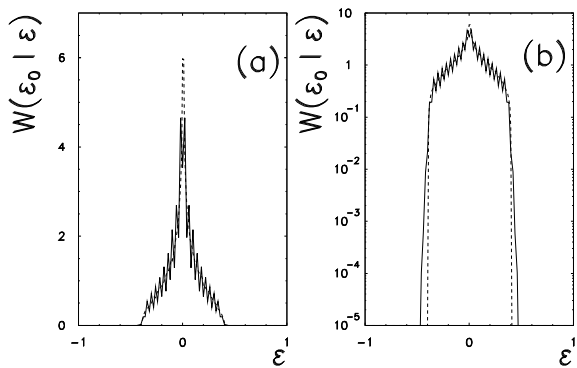}
\vspace{-2cm}
\narrowtext
\caption{a) Shape of eigenfunctions (the $F-$function)
in the energy representation
(full line) and its
classical analog (dashed line) for the parameters of Fig.2,
after [24].}
\label{ef}
\end{figure}

The interesting question arises in the view of the form of the LDOS, see Fig.2. One can suggest that this form is close to the Breit-Wigner dependence (\ref {BW}) which generically occurs in random matrix models. However, the detailed study \cite{BGIC98} rejected this possibility since only in a very small energy region close to the center one can speak about a kind of the correspondence. This means that for the dynamical model of two interacting spins typical form of the LDOS is neither the Breit-Wigner nor the Gaussian. One should note that in this model there are no free parameters (apart from the total energy) which can change the relative strength of the interaction $V$ .

When comparing the shape of eigenstates with the form of the LDOS, see
Figs.2-3, one can see a strong difference in the energy range associated
with the width of energy shells. This fact is due to the difference in the
density of states$\,\rho _0(E)\,$ and $\rho (E)$ for the unperturbed and
total Hamiltonians, see details in \cite{BGIC98,BGI98}. It was, however,
shown that with a proper rescaling the LDOS distribution coincides with the $%
F-$ function with a high accuracy.

\section{One particle in a rippled channel}

Now let us consider the motion of a particle in the quasi-1D rippled
billiard \cite{LMI00}, see Fig. 4.

\begin{figure}
\epsfxsize 7cm
\epsfbox{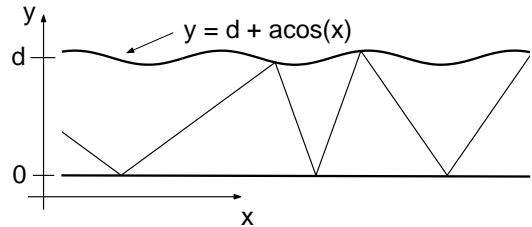}
\narrowtext
\caption{Geometry of the rippled channel,
after [26].
}
\end{figure}

It was shown \cite{LKRH96,LRK96} that the dynamics of a classical particle
in this billiard manifests generic properties of the transition to chaos
with an increase of the amplitude $a$ for a fixed $d$. A possible
experimental realization of this model with {\it finite} length is a
mesoscopic electron wave guide. In \cite{LKRH96} {\it classical} transport
properties of this billiard have been related to its dynamical properties,
thus giving a transport signature of chaos. The analysis of the quantum
motion in the {\it infinitely} long rippled channel with the periodic upper
boundary $y(x+L)=y(x)$ is useful for the understanding of universal features
of electronic band structures of real crystals, propagation in periodic
waveguides, quantum wires and films, see references in \cite{LKRH96,LRK96}.
Detailed study of the energy band structure of the quantum version of this
periodic rippled channel is presented in Ref.\cite{LRK96}.

The control parameter that determines the degree of chaos in the classical
model is $K=\frac{2da}\pi$ , and strong chaos occurs when $K \stackrel{>}{%
\sim }1$. In the quantum description one should solve the Schr\"odinger
equation $\hat H=\frac{\hbar ^2}2(\hat p_x^2+\hat p_y^2)$ with the boundary
conditions $\psi (x,y)=0$ at $y=0$ and $y=d+a\cos x$. Instead, in Ref.\cite
{LKRH96,LRK96} the approach has been used which is based on the transition
to the new coordinates $u=x$, and $v=\frac{yd}{d+acos(x)}$, for which the
boundary conditions become trivial, $\Psi (u,v)=0$ at $v=0,d$. On the other
hand, in the new variables the specific interaction between two degrees of
freedom arises and the Hamiltonian acquires a much more complicated form,

\begin{equation}
\hat H=\frac{\hbar ^2}2g^{-1/4}\hat P_\alpha g^{\alpha \beta }g^{1/2}\hat
P_\beta g^{-1/4},\ \ \alpha ,\beta =u,v, 
\end{equation}
which is simply the kinetic energy expressed in the covariant form. The
momentum is now given by $\hat P_\alpha =-i\hbar [\partial _\alpha +\frac
14\partial _\alpha ln(g)]=-i\hbar g^{-1/4}\partial _\alpha g^{1/4}$, where $%
g^{\alpha \beta }$ is the metric tensor, and $g=Det(g_{\alpha \beta
})=[1+\epsilon \cos u]^2$, $\epsilon \equiv a/d$, (see \cite{LRK96}  for
details). As a result, the original model of one particle in the rippled
billiard is formally transformed into the model of two interacting
particles. In this way, the complexity of the boundary in the original model
is incorporated into the ``interaction potential''.

The solution for the wave function can be represented by expanding the
function $\Phi _k(u,v)$ in a double Fourier series, $\Psi ^\alpha
(u,v;k)=\sum_{m=1}^\infty \sum_{n=-\infty }^\infty C_{mn}^\alpha (k)\phi
_{mn}^k(u,v)\,,$ where $\alpha $ stands for the eigenstate of energy $%
E^\alpha (k)$, and $\phi _{mn}^k(u,v)=\pi ^{-1/2}g^{-1/4}sin(m\pi
v)e^{i(k+n)u}$ and the Bloch number was fixed $k=0.1$ in the numerical
study. The index $m$ refers to the mode (or channel) number and $n$ stands
for the Brillouin zone number. Note that $\phi _{mn}^k$ are the eigenstates
of the unperturbed momenta squared (and, therefore, of the unperturbed
Hamiltonian): $\hat P_v^2\phi _{mn}^k=\hbar ^2(m\pi )^2\phi _{mn}^k$, $\hat
P_u^2\phi _{mn}^k=\hbar ^2(k+n)^2\phi _{mn}^k $. One should remind that each
pair $(m,n)$ is associated with one number $l\,$ which labels the
unperturbed two-particle basis and it is essential that the unperturbed
basis of $H_0$ has to be ordered in increasing energy, $E_l^0(k)=\frac{\hbar
^2}2((n+k)^2+(m\pi )^2)$.

In the classical limit one should perform the coordinate transformation in
the same way as it was done for the quantum model. The classical Hamiltonian
has the form $H=\frac 12g^{\alpha \beta }P_\alpha P_\beta ,$ with the same
boundary conditions in $(u,v)$ coordinates. Expanding this expression, one
can obtain,

\begin{equation}
H=\frac 12\left[ P_u^2-\frac{2\epsilon v\xi _u}{1+\epsilon \xi }P_uP_v+\frac{%
1+(\epsilon v\xi _u)^2}{(1+\epsilon \xi )^2}P_v^2\right] , 
\end{equation}
where $\xi \equiv cosu,\xi _u\equiv d\xi /du$, and $\epsilon \equiv a/d$. It
should be noted that the transformation from the old to new variables is
canonical. In accordance with the quantum model, the separation of the total
Hamiltonian into two parts, $H=H^0+V$ , has to be done as follows, 
\begin{equation}
\label{newH}H^0\equiv \frac 12\left[ P_u^2+P_v^2\right] ;\,\,\,\,V\equiv
H-H^0\,\,. 
\end{equation}

Let us now compare the quantum LDOS and the shape of exact eigenstates with
their classical counterparts which are numerically obtained \cite{LMI00} in
the described above approach, see Fig.5. One can see, that, in general,
there is quite good global correspondence between quantum and classical data
(the role of quantum fluctuations around the envelopes will be discuss
below). The interesting point is that the forms of the LDOS and the SE are
highly non-trivial, if to compare with the Breit-Wigner or Gaussian
distribution. Remarkably, the averaging procedure reveals a three-peak
structure for both classical and quantum quantities.

\begin{figure}
\epsfxsize 7cm
\epsfbox{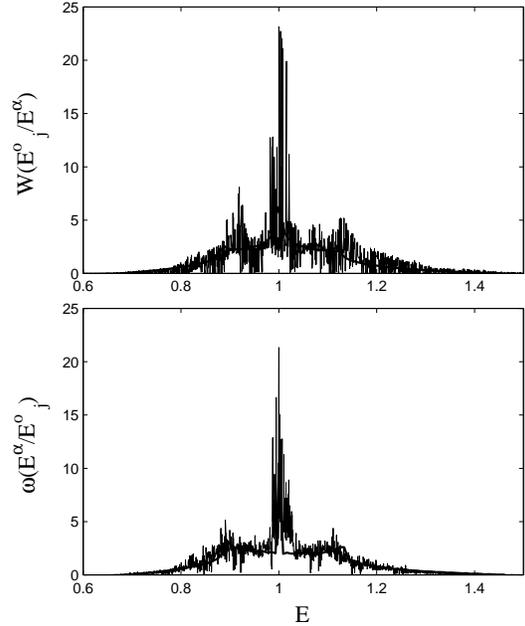}
\narrowtext
\caption{ (above) Shape of eigenfuctions (the $F-$function) in the energy
representation (broken line) and its classical counterpart
(smooth curve); (below) the LDOS and its classical counterpart.
For the $F-$function the average is taken over exact eigenstates from
the interval $1870<\alpha<1950$
and for the LDOS the average is over basis states from
the interval $1900< j <1990$, after [26].
}
\end{figure}

The origin of these peaks can be understood by a detailed analysis of the
classical trajectories \cite{LMI00}. Specifically, it was found that even
for the chaotic motion the particle can be located for a long time in the
neighborhood of stable and unstable periodic orbits of period 1. One can
find that the unstable (stable) periodic orbit is defined by $P_u=0$ and $%
x=0;\pi $ respectively. The right peak corresponds to the motion
perpendicular to the $x$ axis at $x=\pi $ (unstable orbit). Similarly, the
left one results from the stable orbit at $x=0$ . In contrast, the central
peak is originated from the trajectories which are nearly parallel to the $x$%
-axis. A similar analysis explains the origin of the structure of the
classical LDOS .

As for the comparison between the form of the LDOS\ and that of the SE, it
is easy to detect a difference, especially, in the central region. The
reason for this is still not clear, in Ref.\cite{LMI00} it was suggested
that it is due to the difference between time averages and phase space
averages for finite times.

\section{ Localization and non-ergodicity in the energy shell}

Now let us discuss the problem of localization in the energy shell. As was
claimed above, with an increase of the interaction in the TBRI model, the
Anderson-like transition is expected \cite{AGKL97} in the Hilbert space of
many-particle states. It occurs when the interaction strength $V_0$ exceeds
the mean level spacing between basis states which are directly coupled by
the two-body interaction, $V_0\geq d_f$ . In terms of the structure of
compound states, this means that above this threshold the components $%
C_j^{(i)}$ ergodically fill an available energy shell. More specifically,
the fluctuations of these components fluctuate around the mean given by the $%
F-$function, in accordance with the conventional statistical mechanics,
typically, according to the Gaussian distribution. One should stress that
for closed systems the width of the energy shell is always finite due to a
finite range of interaction, and it can be much less than the total size of
the energy spectrum. However, the number $N_{pc}$ of principal components
can be very large. In this sense, one can rigorously define the {\it onset
of chaos} (or maximal complexity in eigenstates), as the ergodic filling of
the energy shell with the Gaussian fluctuations around the envelope given by
the $F-$function.

If the system has the classical limit, the $F-$function has its classical
counterpart which can be easily found from Newton equations of motion. More
difficult situation is when there is no classical limit, in this case,
although the $F-$function exists, it is of the problem to find it without
direct diagonalization (and the average over a number of eigenstates) of a
(typically huge) Hamiltonian matrix. Using two examples of simple quantum
dynamical models with the classical chaos, we have shown here that in a deep
semiclassical limit there is a good correspondence for the form of both the
LDOS and eigenstates. Strange enough, this quantum-classical correspondence
is studied scarcely, there are mainly numerical results discussed above (see
also, \cite{we}). As the data for the rippled channel show, a kind of
semiclassical periodic orbit theory can be developed both for the LDOS\ and
SE.

The knowledge of the SE is extremely important for the study of any kind of
localization in the energy shell. The non-trivial role of the $F-$function
is that it gives a reference to which individual eigenstates should be
compared in order to see the localization. Let us first consider the TBRI
model where there is no influence of classical periodic orbits which result
in a specific sort of localization (scars, if to speak about the
quantum-classical correspondence in the configuration space). For such kind
of models, above the threshold $V_0\approx d_f$ the gaussian fluctuations
for the coefficients $C_j^{(i)}$ are expected. This was checked for excited
states of the Ce atom which is known to be quite chaotic dynamical system 
\cite{C85,FGGK94}. Numerical data have shown, that, indeed, the fluctuations
turns out to be very close to the Gaussian. Note, that for the study of
these fluctuations, first, one needs to normalize each value $C_j^{(i)}$ to
its mean-square-root value which itself depends on the index $i$ (or on the
energy $E^{(i)}$, if to analyze the fluctuations in the energy space).
Therefore, one should find this mean value (square root of the $F-$function)
by a kind of the average.

Below the threshold $V_0\approx d_f$ the fluctuations of the amplitudes $%
C_j^{(i)}$ are extremely large since the components which correspond to
those states, which are not coupled directly by a two--body interaction, are
very small, thus resulting in  a large peak in the distribution of $P(C)\,$
around zero. As one can see, the quantum localization results in the
sparsity of the $F-$function and the LDOS, in contrast to the standard
Anderson localization in one and quasi-1D disordered models, where
eigenstates are ``dense" in the configuration space and occupy a finite
number of sites (analog of basis states in our models). Therefore, the
fingerprint of the Anderson localization is an exponential decay of the
amplitudes $C_k$ far from the ``centers'' of eigenstates, unlike the
localization in the energy shell for closed systems, where the cut-off of
tails always exists due to finiteness of the energy shell. On the other
hand, for a very strong localization, when eigenstates consist of few basis
states (for $V_0\ll d_f$ ), the situation is quite similar to that for the
Anderson localization (this typically happens for low energy states due to a
relatively large values of $d_f$ ) .

A much more delicate situation occurs for dynamical quantum systems with the
classical limit. In this case in addition to the quantum localization, even
in a strongly chaotic classical region, the influence of tiny structure of
the classical phase space (for example, short unstable periodic orbits) may
be quite important. In this sense, one can speak about the localization
which is caused by both classical properties (periodic orbits) and quantum
ones (scarring effect). For this reason, the mesoscopic region where both
classical and quantum effects can be equally strong, is of the most
difficult for the analysis.

Coming back to the numerical data for the rippled channel \cite{LMI00}, a
noticeable fraction of extremely localized eigenstates has been detected in
the region of a strong classical chaos where typically eigenstates are
ergodic. It was found that these eigenstates are originated from those basis
states which are located on the edges of the shells of $H_0$ . It turns out
that these basis states are very slightly coupled with the others by the
interaction $V$ . These localized states correspond to almost free motion of
a particle along the channel. The presence of very localized states has been
also observed in the model (\ref{ham1}) of two interacting spin \cite
{BGIC98,BGI98}. In this case, their origin is not so clear and deserves an
additional study.

Finally, we would like to raise a new problem related to the
quantum-classical correspondence for chaotic classical systems. It refers to
the ergodicity of {\it individual }eigenstates in a deep semiclassical
region. According to the Shnirelman theorem, it is believed that the closer
this limit, the less influence of quantum effects. To see this fact, a kind
of the average over a small energy range is typically assumed. On the other
hand, in a deep semiclassical limit the eigenstates have a very large number
of components. This allows one to perform the statistical analysis inside 
{\it individual eigenstates}, without any energy or ensemble average. This
is in a spirit of dynamical chaos for which statistical properties can be
revealed along a single chaotic trajectory. Specifically, we can ask: how
fast these fluctuations approach those ones prescribed by the standard
statistical mechanics? Numerical data \cite{LMI00} manifest that individual
eigenstates can not be rigorously treated as random ones since
non-statistical deviations seem to persist for {\it any finite values} of $%
\hbar $. In this sense, one can speak about {\it non-ergodic character} of
individual eigenstates, in spite of the existence of the classical limit
itself. Definitely, this question deserves a detailed study, both
numerically and analytically.

\section{Acknowledgments}

The author is very grateful to his co-authors with whom the works have been
done on the subject discussed in this paper. This work was supported by
CONACyT (Mexico) Grant No. 28626-E.

\end{document}